\documentclass[conference]{IEEEtran}
\IEEEoverridecommandlockouts

\usepackage{float} 
\usepackage{hyperref}
\hypersetup{
    colorlinks=true,
    linkcolor=cyan,
    filecolor=black,      
    urlcolor=cyan,
    citecolor=cyan,
}

\usepackage{cite}
\usepackage{amsmath,amssymb,amsfonts}
\usepackage{algorithmic}
\usepackage{graphicx}
\usepackage{textcomp}
\usepackage{xcolor}
\begin{document}

\title{Algorithmic Extremism: Examining YouTube's Rabbit Hole of Radicalization
}

\author{\IEEEauthorblockN{Mark Ledwich}
\IEEEauthorblockA{ 
Brisbane, Australia\\
mark@ledwich.com.au}
\and
\IEEEauthorblockN{Anna Zaitsev}
\IEEEauthorblockA{\textit{The School of Information} \\
\textit{University of California, Berkeley}\\
Berkeley, United States \\
anna.zaitsev@ischool.berkeley.edu}
}

\maketitle

\begin{abstract}

The role that YouTube and its behind-the-scenes recommendation algorithm plays in encouraging online radicalization has been suggested by both journalists and academics alike. This study directly quantifies these claims by examining the role that YouTube’s algorithm plays in suggesting radicalized content. After categorizing nearly 800 political channels, we were able to differentiate between political schemas in order to analyze the algorithm traffic flows out and between each group. After conducting a detailed analysis of recommendations received by each channel type, we refute the popular radicalization claims. To the contrary, these data suggest that YouTube’s recommendation algorithm actively discourages viewers from visiting radicalizing or extremist content. Instead, the algorithm is shown to favor mainstream media and cable news content over independent YouTube channels with slant towards left-leaning or politically neutral channels. Our study thus suggests that YouTube’s recommendation algorithm fails to promote inflammatory or radicalized content, as previously claimed by several outlets. 
\end{abstract}

\begin{IEEEkeywords}
YouTube, recommendation algorithm, radicalization
\end{IEEEkeywords}

\section{Introduction}

The internet can both be a powerful force for good, prosocial behaviors by providing means for civic participation and community organization \cite{ferdinand2013internet}, as well as an attractor for antisocial behaviors that create polarizing extremism \cite{blaya2019cyberhate}. This dual nature of the internet has been evident since the early days of online communication, where "flame-wars" and "trolling" have been present in online communities for over two decades \cite{pfaffenberger1996if}\cite{kayany1998contexts}\cite{berghel2018online}. While such behaviors were previously confined to Usenet message boards and limited IRC channels, with the expansion of social media, blogs, and microblogging following the rapid growth of internet participation rates, these inflammatory behaviors are no longer confined and have left their early back-channels into public consciousness \cite{itu2019stats}. 

The explosion of platforms, as well as ebbs and flows in the political climate, has exacerbated the prevalence of antisocial messaging \cite{gagliardone2015countering}. Research focusing on uninhibited or antisocial communication, as well as extremist messaging online has previously been conducted on platforms including Facebook \cite{ben2016hate}, Twitter \cite{burnap2015cyber}, Reddit \cite{chandrasekharan2017you}, 4chan and 8chan \cite{knuttila2011user}\cite{nagle2017kill}, Tumblr \cite{agarwal2016spider} and even knitting forums such as Ravelry \cite{shen2018perceptions}. 

In addition to these prior studies on other platforms, attention has recently been paid to the role that YouTube may play as a platform for radicalization \cite{roose2019making}\cite{adl2019youtube}\cite{munn2019alt}. As a content host, YouTube provides a great opportunity for broadcasting a large and widely diverse set of ideas to millions of people worldwide. Included among general content creators are those who specifically target users with polarizing and radicalizing political content. While YouTube and other social media platforms have generally taken a strict stance against most inflammatory material on their platform, extremist groups from jihadi terrorist organizations \cite{andre2012neojihadism}\cite{awan2017cyber}, various political positions \cite{de2012social}, and conspiracy theorists have nonetheless been able to permeate the content barrier \cite{schmitt2018counter}. 


Extreme content exists on a spectrum. YouTube and other social media platforms have generally taken a strict stance against the most inflammatory materials or materials that are outright illegal. No social media platform tolerates ISIS beheading videos, child porn, or videos depicting cruelty towards animals. There seems to a consensus amongst all social media platforms that human moderators or moderation algorithms will remove this type of content \cite{gillespie2018custodians}.

YouTube's automatic removal of the most extreme content, such as explicitly violent acts, child pornography, and animal cruelty, has created a new era of algorithmic data mining \cite{agarwal2015topic}\cite{sureka2010mining}\cite{agarwal2016spider}. These methods range from metadata scans \cite{hussain2018analyzing} to sentiment analysis \cite{asghar2015sentiment}. Nevertheless, content within an ideological grey area or that can nonetheless be perceived as "radicalizing" exists on YouTube \cite{youtube2019guidelines}. Definitions of free speech differ from country to country. However, YouTube operates on a global scale within the cultural background of the United States with robust legislation that protects speech \cite{gagliardone2015countering}. Even if there are limitations to what YouTube will broadcast, the platform does allow a fair bit of content that could be deemed as radicalizing, either by accident or by lack of monitoring resources.  

Means such as demonetization, flagging, or comment limiting is several tools available to content moderators on YouTube \cite{youtube2019limited}. Nevertheless, removing or demonetizing videos or channels that present inflammatory content has not curtailed scrutiny of YouTube by popular media \cite{martienau2019hate}. Recently, the New York Times published a series of articles, notably critiquing YouTube's recommendation algorithm, which suggests related videos for users based on their prior preferences and users with similar preferences \cite{tufekci2018youtube}\cite{roose2019making}. The argument put forward by the NYT is that users would not otherwise have stumbled upon extremist content if they were not actively searching for it since the role of recommendation algorithms for content on other websites is less prevalent. As such, YouTube's algorithm may have a role in guiding content, and to some extent, preferences towards more extremist predispositions. Critical to this critique is that while previous comments on the role that social media websites play in spreading radicalization have focused on user contributions, the implications of the recommendation algorithm strictly implicate YouTube's programming as an offender.

The critique of the recommendation algorithm is another difference that sets YouTube apart from other platforms. In most cases, researchers are looking at how the \textit{users} apply social media tools as ways to spread jihadism \cite{andre2012neojihadism}, alt-right messages of white supremacy \cite{nagle2017kill}. Studies are also focusing on the methods the content creators might use to recruit more participants in various movements; for example, radical left-wing Antifa protests \cite{vacca2019online}. Nevertheless, the premise is that users of Facebook, Tumblr, or Twitter would not stumble upon extremists if they are not actively searching for it since the role of recommendation algorithms is less prevalent. There are always some edge cases where innocuous Twitter hashtags can be co-opted for malicious purposes by extremists or trolls \cite{awan2017cyber}, but in general, users get what they specifically seek. However, the case for YouTube is different: the recommendation algorithm is seen as a major factor in how users engage with YouTube content. Thus, the claims about YouTube's role in radicalization are twofold. First, there are content creators that publish content that has the potential to radicalize \cite{roose2019making}. Second, YouTube is being scrutinized for how and where the recommendation algorithm directs the user traffic \cite{munn2019alt}\cite{roose2019making}. Nevertheless, empirical evidence of YouTube's role in radicalization is insufficient \cite{ribeiro2019auditing}. There are anecdotes of a radicalization pipeline and hate group rabbit hole, but academic literature on the topic is scant, as we discuss in the next section.  

\section{Prior Academic Studies on YouTube Radicalization}

Data-drive papers analyzing radicalization trends online are an emerging field of inquiry. To date, few notable studies have examined YouTube's content in relation to radicalization. As discussed, previous studies have concentrated on the content itself and have widely proposed novel means to analyze these data \cite{agarwal2016spider}\cite{agarwal2017metadata}\cite{hussain2018analyzing}. However, these studies focus on introducing means for content analysis, rather than the content analysis itself. 

However, a few studies go beyond content analysis methods. One such study, Ottoni et al. (218), analyzed the language used in right-wing channels compared to their baseline channels. The study concludes that there was little bias against immigrants or members of the LGBT community, but there was limited evidence for prejudice towards Muslims. However, the study did find evidence for the negative language used by channels labeled as right-wing. Nevertheless, this study has a few weaknesses. The authors of this paper frame their analysis as an investigation into right-wing channels but then proceed to analyze kooky conspiracy channels instead of more mainstream right-wing content. They have chosen a conspiracy theorist Alex Jones' InfoWars (nowadays removed from YouTube) as their seed channel, and their list of right-wing channels reflects this particular niche. InfoWars and other conspiracy channels represent only a small segment of right-wing channels. Besides, the study applies a topic analysis method derived from the Implicit Association Test (IAT)\cite{greenwald1998measuring}. However, the validity of IAT has been contested \cite{forscher2019meta}. In conclusion, we consider the seed channel selection as problematic and the range of the comparison channels as too vaguely explained \cite{ottoni2018analyzing}.

In addition to content analysis of YouTube's videos, Riberio et al. (2019) took a novel approach by analyzing the content of video comment sections, explaining which types of videos individual users were likely to comment on overtime. Categorizing videos in four categories, including alt-right, alt-light, the intellectual dark web (IDW), and a final control group, the authors found inconclusive evidence of migration between groups of videos. \footnote{The study borrows a definition for the alt-right from Anti-Defamation League: \textit{"loose segment of the white supremacist movement consisting of individuals who reject mainstream conservatism in favor of politics that embrace racist, anti-Semitic and white supremacist ideology"} (pp. 2 \cite{ribeiro2019auditing}). The alt-light is defined to be a civic nationalist group rather than racial nationalism groups. The third category, "intellectual dark web" (IDW), is defined as a collection of academics and podcasters who engage in controversial topics. The fourth category, the control group, includes a selection of channels form fashion magazine channels such as the (\href{https://www.youtube.com/user/HelloStyleChannel}{Cosmopolitan} and \href{https://www.youtube.com/user/GQVideos}{GQ Magazine}) to a set of left-wing and right-wing mainstream media outlets.}

The analysis shows that a portion of commenters does migrate from IDW videos to the alt-light videos. There is also a tiny portion of commenter migration from the centrist IDW to the potentially radicalizing alt-right videos. However, we believe that one cannot conclude that YouTube is a radicalizing force based on commenter traffic only. There are several flaws in the setting of the study. Even though the study is commendable, it is also omitting the migration from the center to the left-of-center altogether, presenting a somewhat skewed view of the commenter traffic. In addition, only a tiny fraction of YouTube viewers engage in commenting. For example, the most popular video by Jordan Peterson, a central character of the IDW, has 4.7 million views but only ten thousand comments. Besides, commenting on a video does not necessarily mean agreement with the content. A person leaving a comment on a controversial topic might stem from a desire to get a reaction (trolling or flaming) from either the content creator or other viewers \cite{moor2010flaming}\cite{berghel2018online}. We are hesitant to draw any conclusions based on the commenter migration without analyzing the content of the comments. 

The most recent study by Munger and Phillips (2019) directly analyzed YouTube's recommendation algorithm and suggested that the algorithm operated on a simple supply-and-demand principle. That is, rather than algorithms driving viewer preference and further radicalization, further radicalization external to YouTube inspired content creators to produce more radicalized content. The study furthermore failed to find support for radicalization pathways, instead of finding that the growth belonging to the centrist IDW category reflected a deradicalization trend rather than further radicalization. Nevertheless, these authors are critical towards claims that watching content on Youtube will lead to the spread of radical ideas like a "zombie bite" and are further critical of the potential pipeline from moderate, centrist channels to radical right-wing content. 

\section{Analyzing the YouTube Recommendation Algorithm}

Our study focuses on the YouTube recommendation algorithm and the direction of recommendations between different groups of political content. To analyze the common claims from media and other researchers, we have distilled them into specific claims that can be assessed using our data set.

\noindent\textbf{C1 - Radical Bubbles.} Recommendations influence viewers of radical content to watch more similar content than they would otherwise, making it less likely that alternative views are presented. \newline
\textbf{C2 - Right-Wing Advantage.} YouTube's recommendation algorithm prefers right-wing content over other perspectives. \newline
\textbf{C3 - Radicalization Influence.} YouTube's algorithm influences users by exposing them to more extreme content than they would otherwise seek out.\newline
\textbf{C4 - Right-Wing Radicalization Pathway.} YouTube algorithm influences viewers of mainstream and center-left channels by recommending extreme right-wing content, content that aims to disparage left-wing or centrist narratives. \newline

\noindent By analyzing whether the data supports these claims, we will be able to draw preliminary conclusions on the impact of the recommendation algorithm.  

\subsection{YouTube Channel Selection Criteria}

The data for this study is collected from two sources. First, YouTube offers a few tools for software developers and researchers. Our research applies \href{https://developers.google.com/youtube/v3}{an application programming interface (API)} that YouTube provides for other websites that integrate with YouTube and also for research purposes to define the channel information, including view and engagement statistics and countries. However, the YouTube API limited the amount of information we could retrieve and the period it could be kept and was thus not entirely suitable for this study. For this reason, we use an additional scraping algorithm that provides us information on individual video statistics such as views, likes, video title, and closed captions. This algorithm offers data since the first of January, 2018. The scraping algorithm also provides us the primary data applied for this study: the recommendations that YouTube's recommendation algorithm offers for each video. The scraping process runs daily. 

The scraped data, as well as the YouTube API, provides us a view of the recommendations presented to an anonymous account. In other words, the account has not "watched" any videos, retaining the neutral baseline recommendations, described in further detail by YouTube in their recent paper that explains the inner workings of the recommendation algorithm \cite{zhao2019recommending}. One should note that the recommendations list provided to a user who has an account and who is logged into YouTube might differ from the list presented to this anonymous account. However, we do not believe that there is a drastic difference in the behavior of the algorithm. Our confidence in the similarity is due to the description of the algorithm provided by the developers of the YouTube algorithm \cite{zhao2019recommending}. It would seem counter-intuitive for YouTube to apply vastly different criteria for anonymous users and users who are logged into their accounts, especially considering how complex creating such a recommendation algorithm is in the first place. 

\noindent The study includes eight hundred and sixteen (816) channels which fulfill the following criteria: 
\begin{itemize}
    \item Channel has over ten thousand subscribers. 
    \item More than 30 percent of the content on the channel is political. 
\end{itemize}

\noindent The primary channel selection was made based on the number of subscriptions. The YouTube API provides channel details, including the number of subscribers and aggregate views of all time on the channel. The sizes of the bubble are based on the video views in the year 2018m, not the subscriber counts. YouTube also provides detailed info on the views of each video and dislikes, thus providing information on the additional engagement each video receives from the users. 

Generally, only channels that had over ten thousand subscriptions were analyzed. However, if the channel's subscription numbers were lower than our threshold value or there were missing data. However, if the channel is averaging over ten thousand views per month, the channel was still included. 

We based our selection criteria on the assumption that tiny channels with minimal number of views or subscriptions are unlikely to fulfill YouTube's recommendation criteria: "\textit{1) engagement objectives, such as user clicks, and degree of engagement with recommended videos; 2) satisfaction objectives, such as user liking a video on YouTube, and leaving a rating on the recommendation}\cite{zhao2019recommending}."  

Another threshold for the channels was the focus of the content: only channels where more than 30 percent of the content was on US political or cultural news or cultural commentary, were selected. We based the cultural commentary selection on a list of social issues on the website \href{https://www.isidewith.com/}{ISideWith}.  A variety of qualitative techniques compiled the list of these channels. 

The lists provided by Ad Fontes Media provides a starting point for the more mainstream and well-known alternative sites. Several blogs and other websites further list political channels or provide tools for advanced searches based on topics \cite{crawler2019youtube}\cite{socialblade2019youtube}\cite{political2019youtube}. We also analyzed the recent academic studies and their lists of channels such as Ribero et al. (2019) and Munger and Philips (2019). However, not all channels included in these two studies fit our selection criteria. Thus one can observe differences between the channel lists and categories between our research and other recent studies on a similar subject. 

We added emerging channels by following the YouTube recommendation algorithm, which suggests similar content and which fit the criteria and passed our threshold. We can conceptualize the recommendation algorithm as a type of snowball sampling, a common technique applied in social sciences when one is conducting interview-based data collection but also in the analysis of social networks. Each source is "requested" to nominate a few candidates that would be of interest to the study. The researcher follows there recommendations until the informants reveal no new information or the inclusion criteria are met (e.g., channels become too marginal, or content is not political). In our case, there is a starting point; a channel acts as a node in the network. Each connected channel (e.g., node) in the network is visited. Depending on the content of the channel, it is either added to the collection of channels or discarded. Channels are visited until there are no new channels, or the new channels do not fit the original selection criteria \cite{lee2006statistical}.

\subsection{The Categorization Process}

The categorization of YouTube channels was a non-trivial task. Activist organizations provide lists and classifications, but many of them are unreliable. For example, there are several controversies around the lists of hate groups discussed by the Southern Poverty Law Center (SPLC)\cite{thiessen2018SPLC}. Also, there seems to be a somewhat contentious relationship between the Anti-Defamation League and YouTubers \cite{mandel2017ald}\cite{alexander2019adl}. We decided to create our categorization, based on multiple existing sources.

First, one has several resources to categorize mainstream or alternative media outlets. Mainstream media such as CNN or Fox News have been studied and categorized over time by various outlets \cite{eberl2017one} \cite{ribeiro2018media}. In our study, we applied two sites that provide information on the political views of mainstream media outlets: \href{https://www.adfontesmedia.com}{Ad Fontes Media} and  \href{https://mediabiasfactcheck.com/}{Media Bias Factcheck}. Neither website is guaranteed to be unbiased, but by cross-referencing both, one can come to a relatively reliable categorization on the political bias of the major news networks. These sites covered the fifty largest mainstream channels, which make up for almost 80 percent of all YouTube views. 

Nevertheless, the majority of the political YouTube channels were not included in sources categorizing mainstream outlets. After reviewing the existing literature on political YouTube and the categorization created by authors such as Ribero et al. (2019) or Munger and Philips (2019), we decided to create a new categorization. Our study strives for a granular and precise classification to facilitate a deep dive into the political subcultures of YouTube, and the extant categories were too narrow in their scope. We decided to apply on both a high-level left-center-right political classification for high-level analysis and create a more granular distinction between eighteen separate labels, described shortly in Table \ref{shorttags} or at length in Appendix \ref{soft}). 

\begin{table}[ht]
\caption{Categorization Soft Tags and Examples}
\begin{tabular}{|p{6cm}|p{2cm}|}
\hline
\textbf{Tag} & \textbf{Examples} \\
\hline
\textbf{Conspiracy}
A channel that regularly promotes a variety of conspiracy theories. &  
X22Report, The Next News Network\\
\hline
\textbf{Libertarian}
Political philosophy with liberty as the main principle. 
&
Reason, John Stossel, The Cato Institute\\
\hline
\textbf{Anti-SJW}
Have a significant focus on criticizing "Social Justice" (see next category) with a positive view of the marketplace of ideas and discussing controversial topics.& 
Sargon of Akkad, Tim Pool\\
\hline
\textbf{Social Justice}
Promotes identity Politics and intersectionality
& 
Peter Coffin, hbomberguy\\
\hline
\textbf{White Identitarian}
Identifies-with/is-proud-of the superiority of "whites" and western civilization.& 
NPIRADIX (Richard Spencer)\\
\hline
\textbf{Educational}
Channel that mainly focuses on education material. &
TED, SoulPancake\\
\hline
\textbf{Late Night Talk shows}
Channel with content presented humorous monologues about the daily news. &
Last Week Tonight, Trevor Noah\\
\hline
\textbf{Partisan Left}
Focused on politics and exclusively critical of Republicans. &
The Young Turks, CNN\\
\hline
\textbf{Partisan Right}
Channel mainly focused on politics and exclusively critical of Democrats, supporting Trump. &
Fox News, Candace Owens\\
\hline
\textbf{Anti-theist}
Self-identified atheist who are also actively critical of religion. &
CosmicSkeptic, Matt Dillahunty\\
\hline
\textbf{Religious Conservative}
A channel with a focus on promoting Christianity or Judaism in the context of politics and culture. &
Ben Shapiro, PragerU\\
\hline
\textbf{Socialist (Anti-Capitalist)}
Focus on the problems of capitalism. 
&
Richald Wolf, NonCompete\\
\hline  
\textbf{Revolutionary}
Endorses the overthrow of the current political system.&
Libertarian Socialist Rants, Jason Unruhe\\
\hline
\textbf{Provocateur}
Enjoys offending and receiving any kind of attention.&
StevenCrowder, MILO\\
\hline
\textbf{MRA (Men’s Rights Activist)}
Focus on advocating for rights for men.&
Karen Straughan\\
\hline
\textbf{Missing Link Media}
Channels not large enough to be considered "mainstream." &
Vox, NowThis News\\
\hline
\textbf{State Funded}
Channels funded by governments. &
PBS NewsHour, Al Jazeera, RT\\
\hline
\textbf{Anti-Whiteness}
A subset of Social Justice that in addition to intersectional beliefs about race &
African Diaspora News Channel\\
\hline  
\end{tabular}
\label{shorttags}
\end{table}

In addition to these 'soft tags,' we applied a set of so-called 'hard tags.' These additional tags allowed us to differentiate between YouTube channels that were part of mainstream media outlets and independent YouTubers. The hard tags are discussed in more detail in Appendix \ref{categories}. The difference between 'soft' and 'hard' tags is that hard tags were based on external sources, whereas the soft tags were based on the content analysis of the labelers. 

The tagging process allowed each channel to be characterized by a maximum of four different tags to create meaningful and fair categories for the content. In addition to labeling created by the two authors, we recruited an additional volunteer labeler, who was well versed in the YouTube political sphere, and whom we trusted to label channels by their existing content accurately. When two or more labelers defined a channel by the same label, that label was assigned to the channel. When the labelers disagreed and ended in a draw situation, the tag was not assigned. The majority was needed for a tag to be applied. 

\begin{figure}[ht]
\centering
    \includegraphics[width=\linewidth]{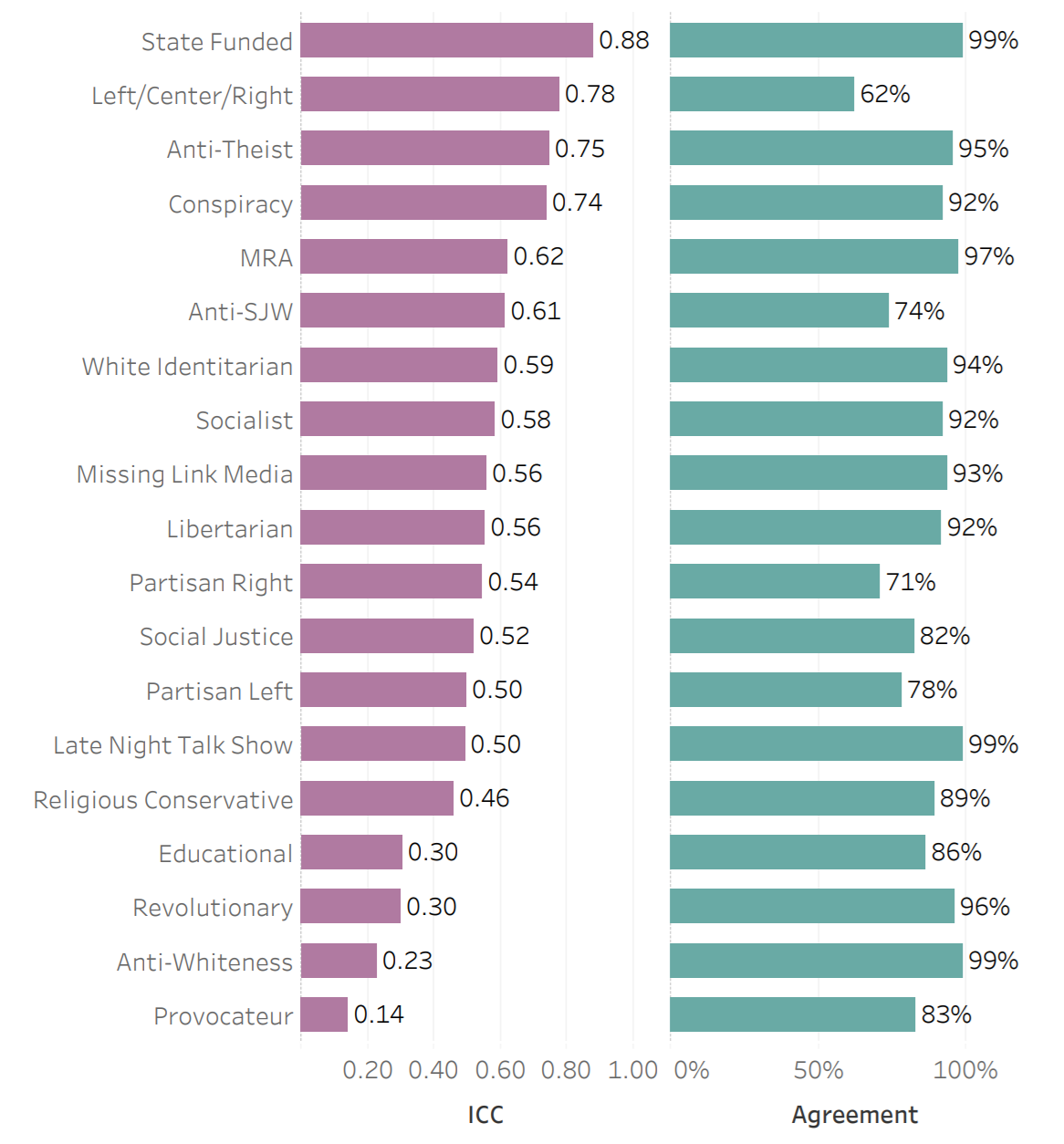}
    \caption{The intraclass correlation coefficiency between the three labelers}
    \label{agreement}
\end{figure}

The visual analysis in Figure \ref{agreement} shows the intraclass correlation coefficiency (ICC) between the three labelers. Based on this analysis, we can determine that all three labelers were in agreement when it comes to the high-level labels, e.g., left-right-center. Besides, there is a high coefficiency in the majority of the granular categories. On the left side of the graph, we can see the intraclass correlation coefficiency values, the estimates of the "real" information captured by our classification, which ranges from 0 to one. The larger the number, the more similar the tags were. One the right side of the Figure, we see the reviewer agreement in percentages. 

The ICC values above 0.75 are considered excellent, between 0.75 and 0.59 are good and above 0.4 are considered as fair \cite{cicchetti1994guidelines}. In our categorization, few classifications measure under 0.4. However, we believe that the explanation for this convergence is related to the nature of these categories. The low coefficiency scoring of groups,'Provocateur', 'Anti-whiteness' and ''Revolutionary,' could be explained by the labeler's hesitation to apply these rather extreme labels where consistent evidence was lacking. Besides, since each channel was allowed four different 'soft tags' defining these subcategories, the channels were likely tagged by the other, milder tags. The rationale behind the lack of agreement on the 'Educational' label is best explained by the fact that this category classification might be somewhat superfluous. Political content, even educational one, often has a clear bias, and the content already belongs to one or more stronger categories, such as Partisan Left or to channels that are non-political.  

However, if one looks at the percentages of the agreement, the agreement if very high in most cases. The only category where disagreement seems to be significant is the left-right-center categorization. However, this disagreement can be explained by the weighing applied when calculating the ICC factor.

To assign a label, we investigated which topics the channels discussed and from which perspective. Some channels are overtly partisan or declare their political stances and support for political parties in their introductions or have posted several videos where such topics are discussed. For example, libertarian channels support Ron and Rand Paul (Libertarian politicians affiliated with the Republican party) or discuss Austrian economics with references to economists such as Frederick von Hayek or Ludwig von Mises or the fictional works of the author Ayn Rand. Comparably, many channels dedicated to various social justice issues title their videos to reflect the content and the political slant, e.g., \href{https://www.youtube.com/watch?v=Q7zDoLTux0Y}{"Can Our Planet Survive Capitalism"} or \href{https://www.youtube.com/watch?v=Pgm9s0ze438}{"The Poor Go To Jail And The Rich Make Bail In America"} from \href{https://www.youtube.com/channel/UCV3Nm3T-XAgVhKH9jT0ViRg}{AJ+}. 

Nevertheless, other channels are more subtle and required more effort to tease out their affiliation. In these cases, we analyzed the perspective that these channels took on political events that have elicited polarized opinions (for example, the nomination of Brett Kavanaugh in the U.S. Supreme Court, the Migrant Caravan, Russiagate). Similarly, we also analyzed the reactions that the channels had for polarizing cultural events or topics (e.g., protests at university campuses, trans activism, free speech). If the majority of these considerations aligned in the same direction, then the channel was designated as left-leaning or right-leaning. If there was a mix, then the channels were likely assigned to the centrist category. 

The only way to conduct this labeling was to watch the content on the channels until the labelers found enough evidence for assigning specific labels. For some channels, this was relatively straightforward: the channels had introductory videos that stated their political perspectives. Some of the intros are very clearly indicating the political viewers of the content creator; some are more subtle. For example, a political commentator Kyle Kulinski explicitly states his political leanings (libertarian-left) in channel \href{https://www.youtube.com/user/SecularTalk}{SecularTalk} description. In contrast, a self-described Classical Liberal discussion host Dave Rubin has a short introduction of various guests, providing examples of the political discussions that take place on his channel \href{https://www.youtube.com/user/RubinReport}{The Rubin Report}. In other cases, the labelers could not assign a label based on introduction or description but had to watch several videos on the channel to determine the political leanings. On average, every labeler watched over 60 hours of YouTube videos to define the political leanings without miscategorizing the channel and thus misrepresenting the views of the content creators. 

Based on the eighteen classification categories, we created thirteen aggregate groups that broadly represent the political views of the YouTube channels. The eighteen 'soft tags' were aggregated from ideological groups and better differentiated between the channels. For more details on tagging aggregation, please see the Appendix \ref{aggregation}. These groupings were applied in the data visualization rather than the more granular eighteen categories for clarity and differentiation purposes. The next section will discuss the data in more detail.   

\section{Findings and Discussion}
\label{findings}

The data on YouTube channels a viewership each channel garners provides us with insights as to how the recommendation algorithm operates. 

\begin{figure}[ht]
\centering
    \includegraphics[width=\linewidth]{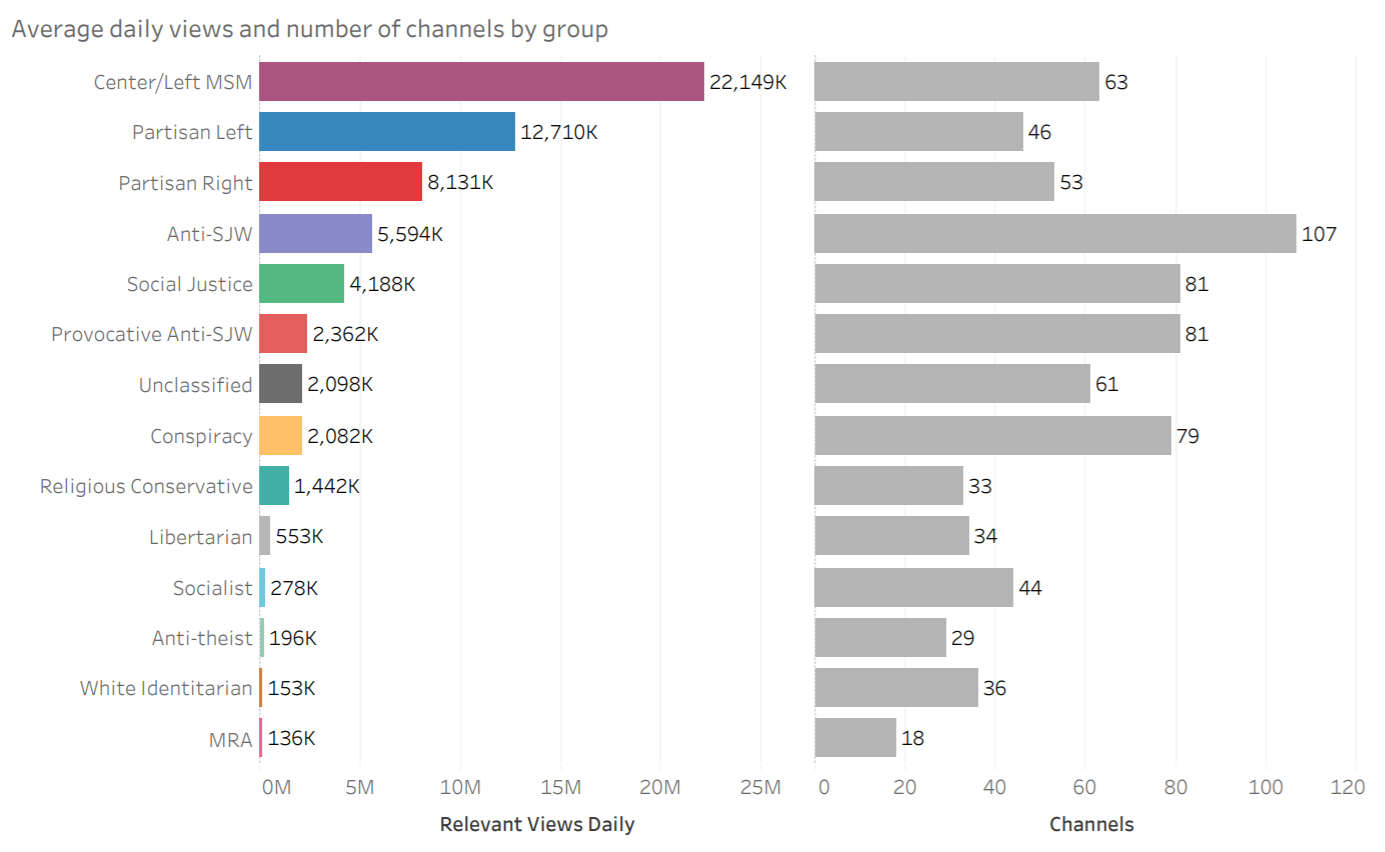}
    \caption{Daily Views and Number of Channels}
    \label{numbers2}
\end{figure}

Per the data collected for 2019, YouTube hosted more channels with content that could be considered right-wing than before. In defining right-wing, we considered categories such as proactive "Anti-SJW" (for anti-Social Just Warrior, a term describing feminist/intersectionality advocates), Partisan-Right, Religious Conservative, and to some extent Conspiracy Channels (for brief explanations, see Table \ref{shorttags}). For longer descriptions on the labels, see Appendix \ref{categories}). However, these more numerous channels gained only a fraction of the views of mainstream media and centrist channels. Categories such as the Center/Left MSM category, Unclassified category (consisting mainly of centrist, non-political and educational channels), and Partisan Left, capture the majority of viewership. The difference here is considerable: where Center/left MSM has 22 million daily views, the largest non-mainstream category, Anti-SJW, has 5.6 million daily views. Figure \ref{numbers2} illustrates the number of views for each category compared to the number of channels. 

\footnote{The Figure \ref{numbers2} and all the following Figures are applying the aggregated categories rather than the granular labels show in Figure \ref{agreement} and discussed in Appendix \ref{aggregation}.} 

\begin{figure}[ht]
\centering
    \includegraphics[width=\linewidth]{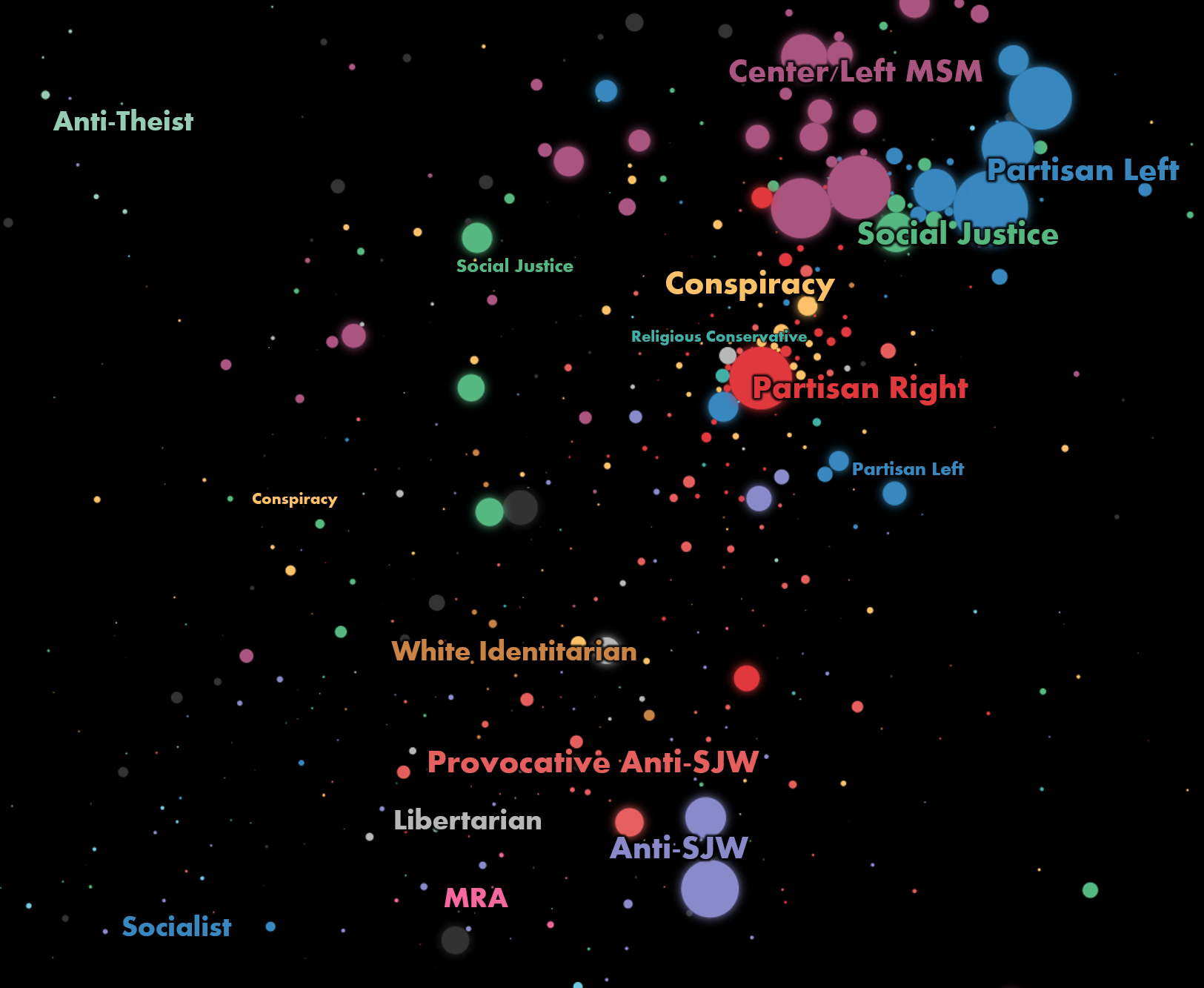}
    \caption{Channel Clusters}
    \label{bubblechart}
\end{figure}

Figure \ref{bubblechart}  presents a chart of channel relations illustrating relations between channels and channel clusters based on the concept of a force-directed graph \cite{bannister2012force}. The area of each bubble, but not the radius, corresponds to the number of views a channel has. The force/size of the line links between channels corresponds to the portion of recommendations between these channels. From this chart, we can see left-wing and centrist mainstream media channels are clustered tightly together. The Partisan Right cluster is also closer to the large mainstream media cluster than it is to any other category. Anti-SJW and Provocative Anti-SJW are clustered tightly together with libertarian channels, while smaller categories such as Anti-theists and socialists are very loosely linked to a limited number of other categories. White Identitarian channels are small and dispersed across the graph.

When analyzing the recommendation algorithm, we are looking at the \textit{impressions} the recommendation algorithm provides viewers of each channel. 

By impressions, we are referring to an estimate for the number of times a viewer was presented with a specific recommendation. This number is an estimate because only YouTube is privy to the data reflecting granulated impressions. However, public-facing data obtained from channels themselves provide us with information on at least the top ten recommendations. A simplified formula for calculating the number of impressions from Channel A to Channel B is calculated by dividing the number of recommendations from A to B by the number of total recommendations channel A receives summed with channel views and recommendations per video, multiplied by ten (for further information, see Appendix \ref{formulas}). Such a calculation of impressions allows us to aggregate the data between channels and categories.

\begin{figure}[ht]
\centering
    \includegraphics[width=\linewidth]{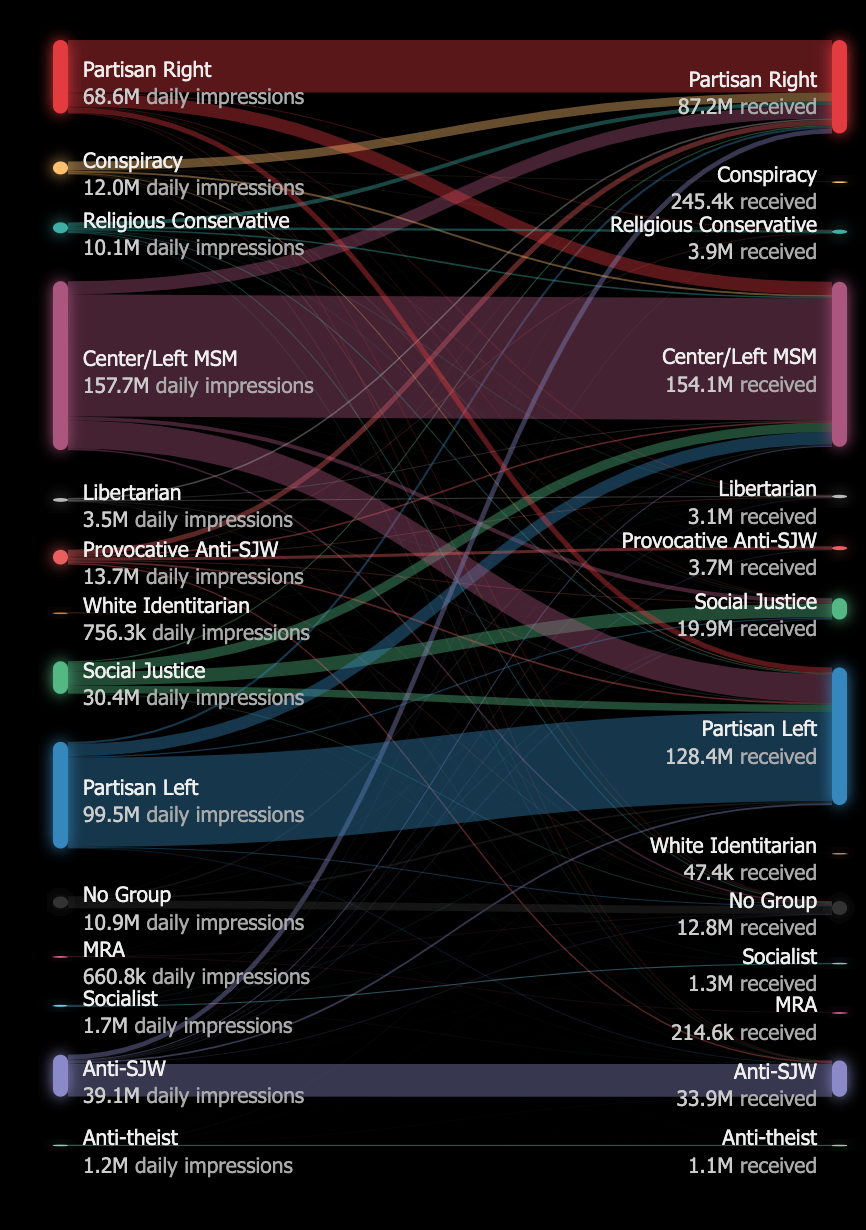}
    \caption{Flow diagram presenting the flow or recommendations between different groups}
    \label{flow}
\end{figure}

Figure \ref{flow} presents the recommendation algorithm in a flow diagram format. The diagram shows the seed channel categories on the left side and the recommendation channel categories on the left side. The sizes of channel categories are based on overall channel view counts. The fourth category from the top is the most viewed channel category, the Center/Left Mainstream media category (MSM). This group is composed of late-night talk shows, mainstream media shows, including the New York Times' YouTube channel. The Partisan Left category closely follows the Center/Left MSM category, with the primary differentiating factor being that the Partisan Left category includes the content of independent YouTube creators. Together, these two most viewed categories garner close to forty million daily views.


Several smaller categories follow the top two-categories. Notably, the two-second largest categories are also centrist or left-leaning in their political outlook. For example, the two largest channels in the Anti-SJW category (\href{https://www.youtube.com/channel/UCnxGkOGNMqQEUMvroOWps6Q}{JRE Clips} and \href{https://www.youtube.com/user/PowerfulJRE/}{PowerfulJRE})) both belong to an American podcast host, Joe Rogan, who hosts guests from a wide range of political beliefs. The Unclassified groups consist of centrist, mostly apolitical, educational channels such as TED or government-owned mainstream media channels such as Russia Today. Based on our flow diagram, we can see that the recommendation algorithm directs traffics from all channel groups into the two largest ones, away from more niche categories.

Based on these data, we can now evaluate the claims that the YouTube recommendation algorithm will recommend content that contributes to the radicalization of YouTube's user base. By analyzing each radicalization claim and whether the data support these claims, we can also conclude whether the YouTube algorithm has a role in political radicalization.

The first claim tested is that YouTube creates \textit{\textbf{C1 - Radical Bubbles.}, i.e., recommendations influence viewers of radical content to watch more similar content than they would otherwise, making it less likely that alternative views are presented.} Based on our data analysis, this claim is partially supported. The flow diagram presented in Figure \ref{flow} shows a high-level view of the intra-category recommendations. The recommendations provided by the algorithm remain within the same category or categories that bear similarity to the original content viewed by the audience. However, from the flow diagram, one can observe that many channels receive fewer impressions than what their views are i.e., the recommendation algorithm directs traffic towards other channel categories. A detailed breakdown of intra-category and cross-category recommendations is presented by recommendations percentages in Figure \ref{advantagepercent} and by a number of impressions in Figure \ref{advantagenumber} in Appendix \ref{advantagedisadvantage} show the strength of intra-category recommendations by channel.  

We can see that the recommendation algorithm does have an intra-category preference, but this preference is dependent on the channel category. For example, 51 percent of traffic from Center Left/MSM channels is directed to other channels belonging to the same category (see Figure \ref{advantagepercent}). Also, the remaining recommendations are directed mainly to two categories: Partisan Left (18.2 percent) and Partisan Right (11 percent), both primarily consisting of mainstream media channels.

\begin{figure}[ht]
\centering
    \includegraphics[width=\linewidth]{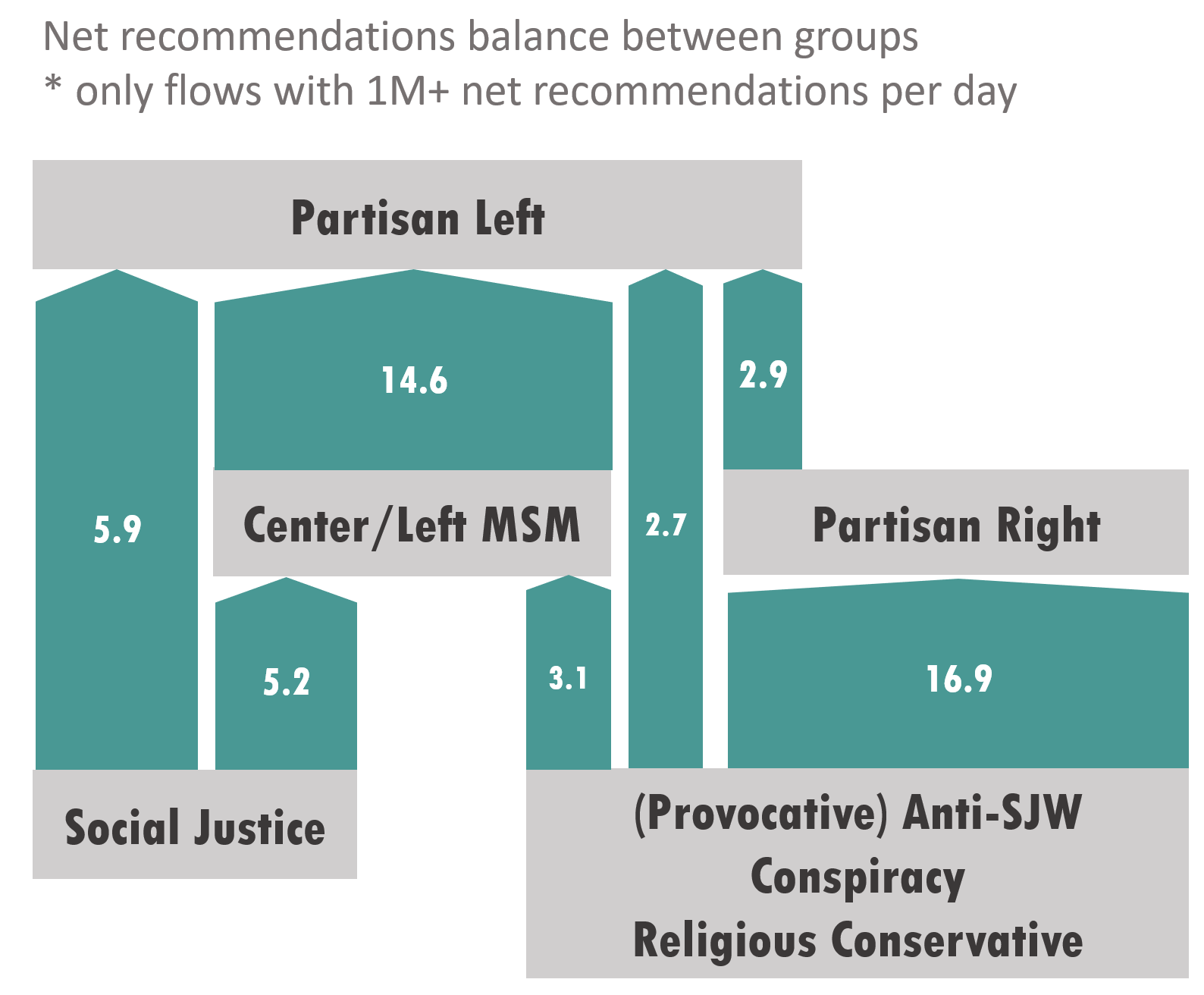}
    \caption{The Direction of Algorithmic Recommendations}
    \label{trafficdirect}
\end{figure}

Figure \ref{trafficdirect} presents a simplified version of the recommendation flows, highlighting the channel categories that benefit from the recommendations traffic. From this figure, we can observe that there is a significant net flow of recommendations towards channels that belong to the category Partisan Left. For example, the Social Justice category suffers from cross-category recommendations. For viewers of channels that are categorized as Social Justice, the algorithm presents 5.9 more recommendations towards the Partisan Left channels than vice versa and another 5.2 million views per day towards Center/Left MSM channels. Figure \ref{trafficdirect} also shows a "pipeline" that directs traffic towards the Partisan Left category from other groups via the intermediary Center/Left MSM category. This is true even for the other beneficiary category, the Partisan Right, which loses 2.9 million recommendations to Partisan Left but benefits with a net flow of recommendations from different right-leaning categories (16.9M).

However, when it comes to categories that could be potentially radicalizing, this statement is only partially supported. Channels that we grouped into Conspiracy Theory or White Identitarian have very low percentages of recommendations within the group itself (as shown in \ref{advantagepercent}). In contrast, channels that we categorized into Center/Left MSM or Partisan Left or Right have higher numbers for recommendations that remain within the group.  These data show that a dramatic shift to more extreme content, as suggested by media \cite{roose2019making}\cite{tufekci2018youtube}, is untenable.

Second, we posited that there is a \textit{\textbf{C2 - Right-Wing Advantage}}, i.e., YouTube's recommendation algorithm prefers right-wing content over other perspectives. This claim is also not supported by the data. On the contrary, the recommendation algorithm favors content that falls within mainstream media groupings. YouTube has stated that its recommendations are based on content that individual users watch and engage in and that peoples' watching habits influence 70 percent of recommendations.

\begin{figure}[ht]
\centering
    \includegraphics[width=\linewidth]{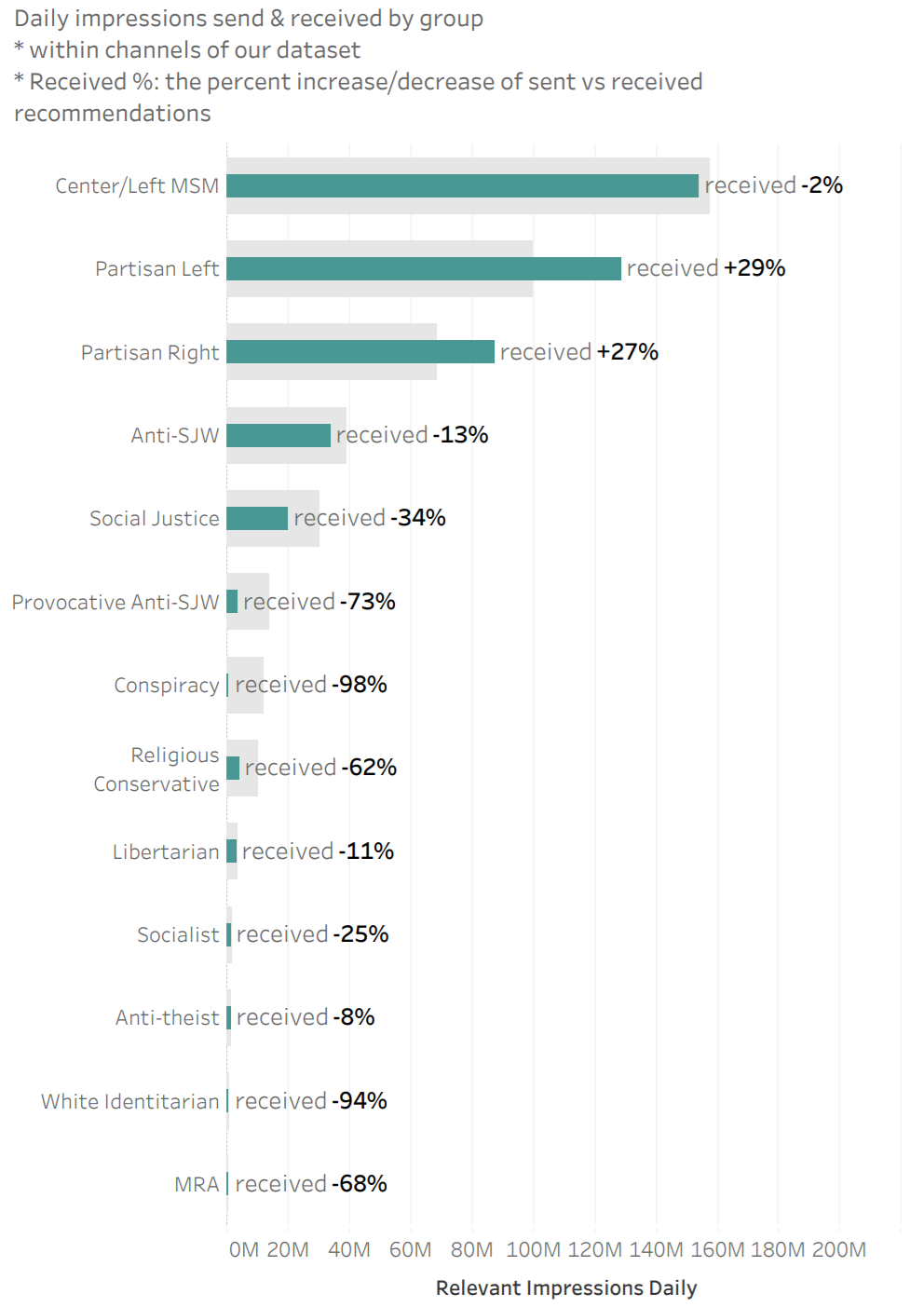}
    \caption{Algorithmic Advantage by Groups}
    \label{advantage}
\end{figure}

Figure \ref{advantage} shows the algorithmic advantage based on daily views. From this Figure, we can observe that the two out of the top three categories (Partisan Left, and Partisan Right) receive more recommendations than other categories irregardless of what category the seed channels belong to. Conversely, any other category does not get their channels suggested by the algorithm. In other words, the recommendation algorithm influences the traffic from all channels towards Partisan Left and Partisan Right channels, regardless of what category the channel that the users viewed belonged to. 

\begin{figure}[ht]
\centering
    \includegraphics[width=\linewidth]{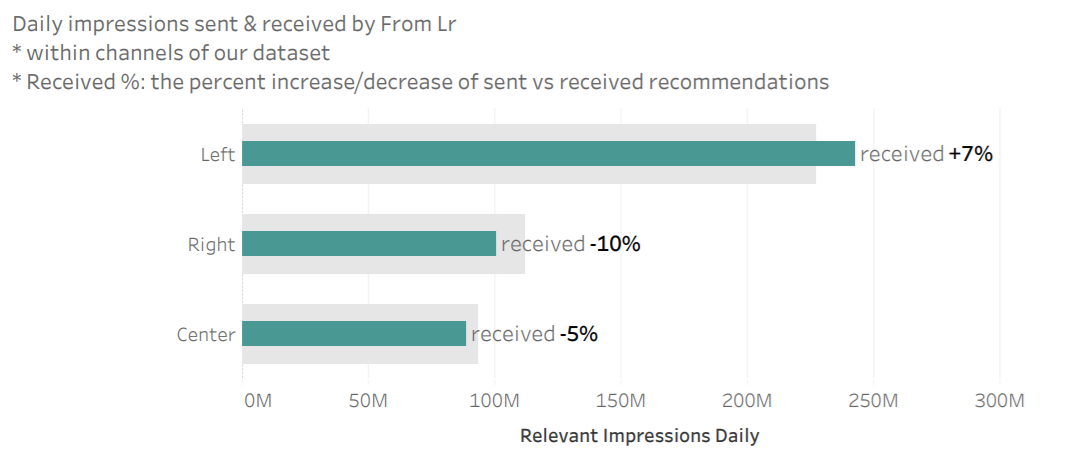}
    \caption{High-level view of Algorithmic Advantages/Disadvantages in Recommendation Impressions}
    \label{highleveladvantage}
\end{figure}

We can also observe this trend from a higher-level aggregate categorization, as is presented in Figure \ref{highleveladvantage}. The Figure affirms that channels that present left or centrist political content are advantaged by the recommendation algorithm, while channels that present content on the right are at a disadvantage.

The recommendations algorithm advantages several groups to a significant extent. For example, we can see that when one watches a video that belongs to the Partisan Left category, the algorithm will present an estimated 3.4M impressions to the Center/Left MSM category more than it does the other way. On the contrary, we can see that the channels that suffer the most substantial disadvantages are again channels that fall outside mainstream media. Both right-wing and left-wing YouTuber channels are disadvantaged, with White Identitarian and Conspiracy channels being the least advantaged by the algorithm. For viewers of conspiracy channel videos, there are 5.5 million more recommendations to Partisan Right videos than vice versa. 

We should also note that right-wing videos are not the only disadvantaged groups. Channels discussing topics such as social justice or socialist view are disadvantaged by the recommendations algorithm as well. The common feature of disadvantages channels is that their content creators are seldom broadcasting networks or mainstream journals. These channels are independent content creators. 

When it comes to the third claim regarding YouTube's potential  \textit{\textbf{C3 - Radicalization Influence}, i.e., YouTube's algorithm influences users by exposing them to more extreme content than they would otherwise}; this claim is also not supported by our data. On the contrary, the recommendation algorithm appears to restrict traffic towards extreme right-wing categories actively. The two most drastic examples are channels we have grouped under the categories of White Identitarian and Conspiracy theory channels. These two groups receive almost no traffic based on the recommendation algorithm, as presented in Figures \ref{advantagepercent} and \ref{advantage}. 

\begin{figure}[ht]
\centering

\includegraphics[width=\linewidth]{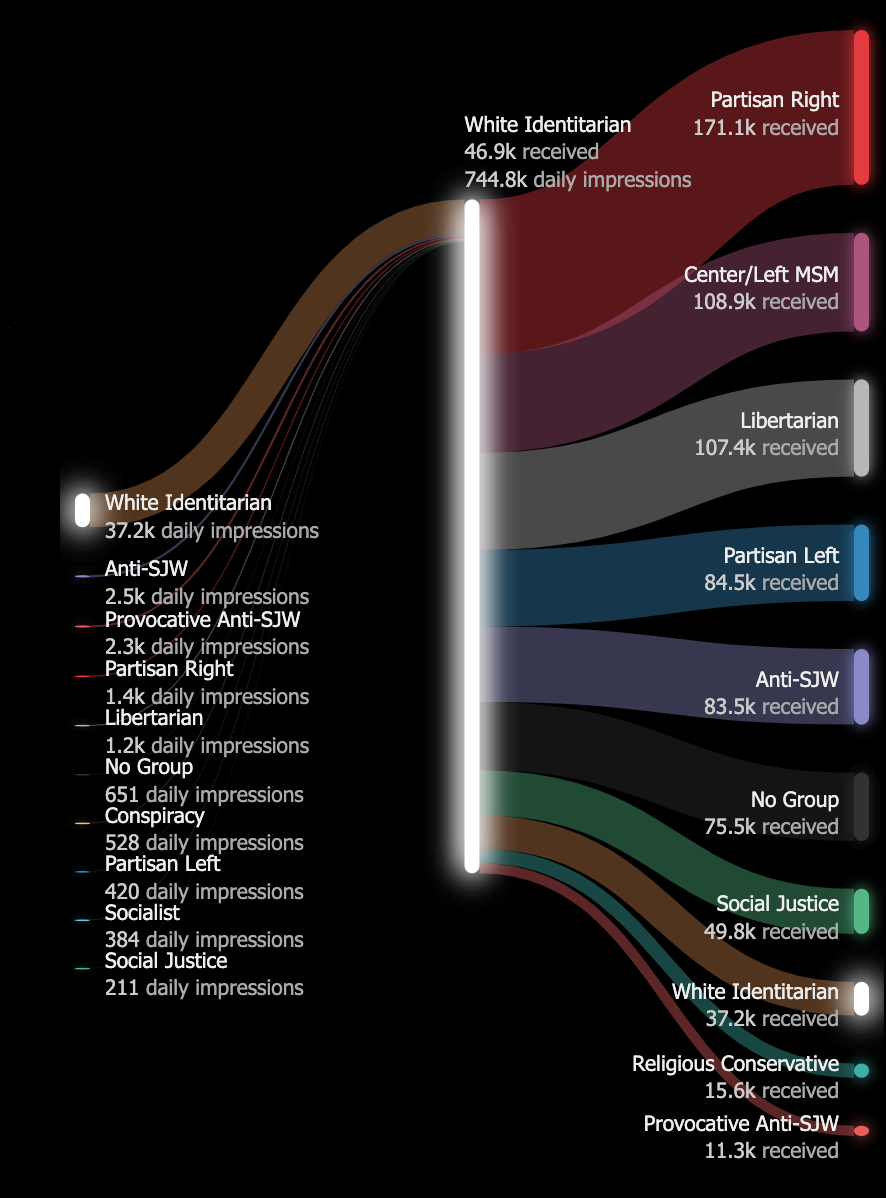} 
\caption{Traffic from White Identitarian Channels}
 \label{WhiteID}
\end{figure}

 \begin{figure}[ht]
\includegraphics[width=\linewidth]{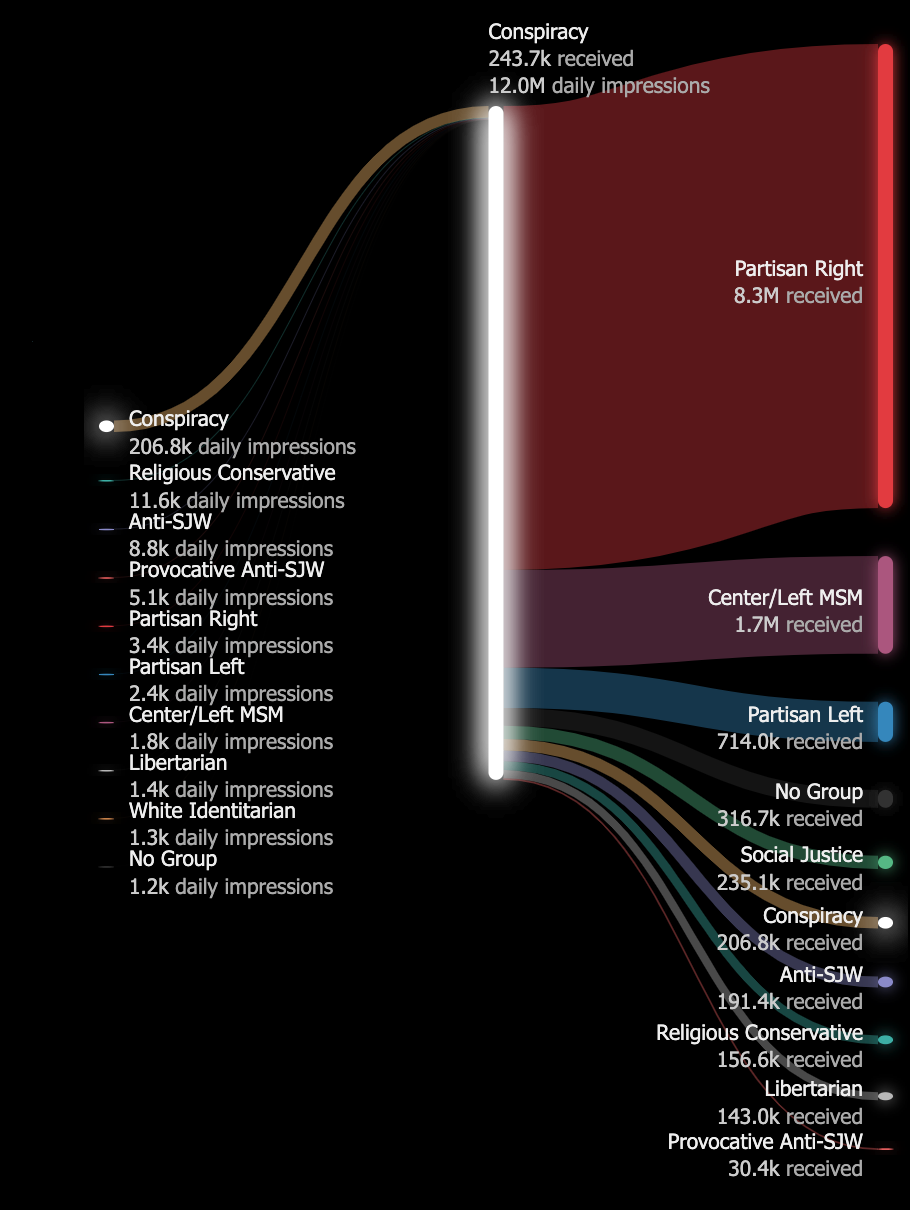}
\caption{Traffic from Conspiracy Channels}
    \label{Consp}
\end{figure}

Another way to visualize the lack of traffic from recommendations is to view the recommendations' flow. Figures \ref{WhiteID} and \ref{Consp} show that the majority of the recommendations flow to either towards Partisan Right, Center/Left MSM, and Partisan Left content. The White Identitarian channel traffic is also directed towards Libertarian and, to a small extent, even towards centrist Anti-SJW content.   

Besides, the Figure \ref{numbers2} showed that the daily views for White Identitarian channels are marginal. Even if we would compare the views of White Identitarian channels with the Conspiracy channels, we could see that Conspiracy channels are twice as viewed than content created by the White Identitarians. This discrepancy is notable since Conspiracy channels seem to gain zero traffic from recommendations (as shown in Figure \ref{advantagepercent}) and are the least advantaged group of all categories. While MRA (Men's Rights Activists) channels form the smallest category in our study, the White Identitarian category is in the bottom five of all groups. Another comparison that illustrates the marginality of White Identitarian channels is the fact that this group consists of thirty-seven channels with enough views to fit within the scope of the study. The White Identitarian category includes almost the same number of channels as Libertarian channels but receives only a third as many views.

\begin{figure}[ht]
\centering
    \includegraphics[width=\linewidth]{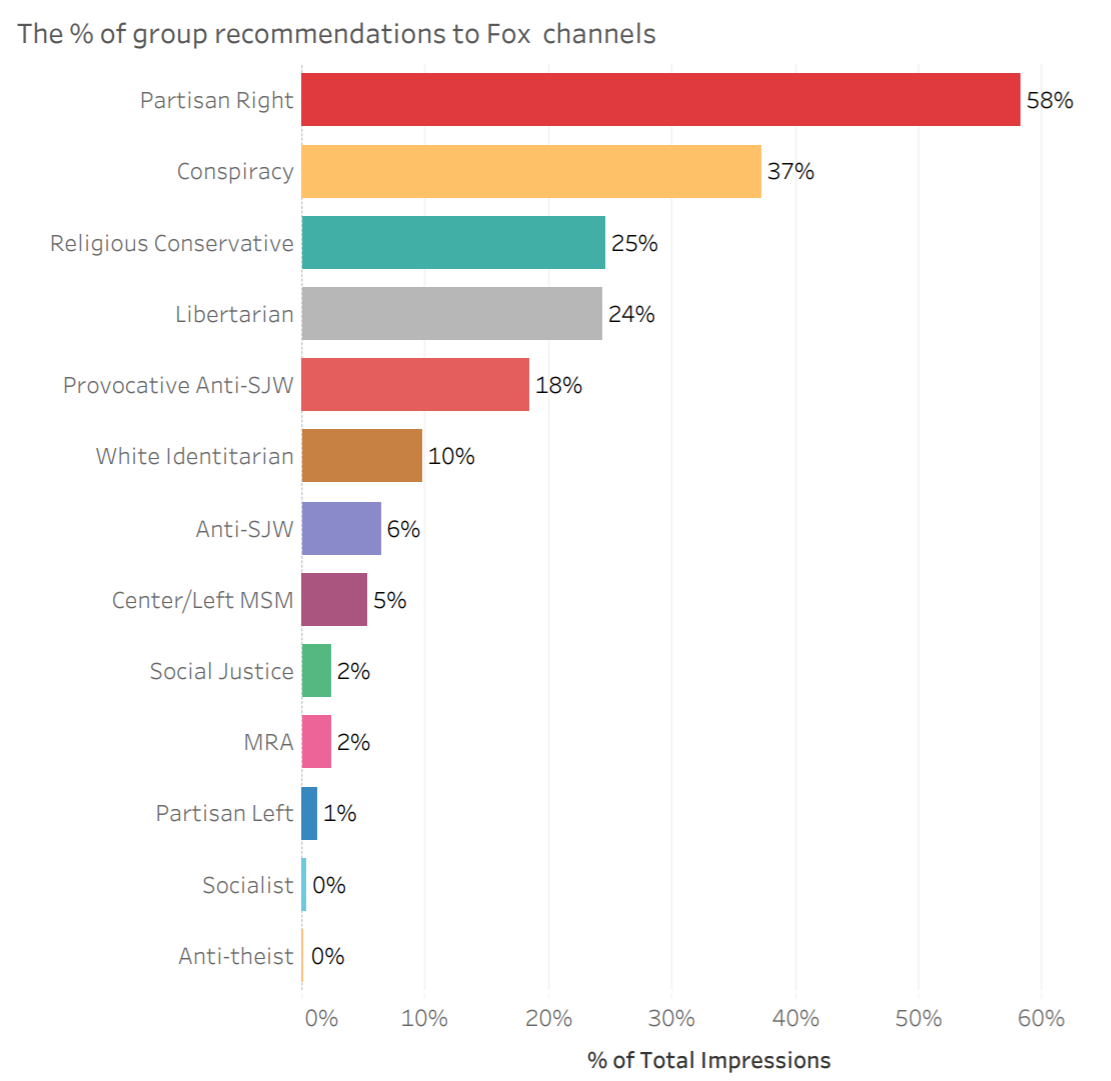}
    \caption{Algorithmic advantage for Fox News}
    \label{fox}
\end{figure}

Our fourth claim stated that there exists \textit{\textbf{C4 - Right-Wing Radicalization Pathway} i.e., YouTube algorithm influences viewers of mainstream and center-left channels via increasingly left-wing critical content to the extreme right.}" Again, these data suggest the opposite. The right-wing channel that benefits the most from the recommendation algorithm is Fox News, a mainstream right-wing media outlet. Figure \ref{fox} shows that Fox News receives over 50 percent of the recommendations form other channels, which map to the category of the Partisan Right. Fox News also receives large numbers of recommendations from every other category that could be considered right-wing. This observation is aligned with the overall trend of the algorithm to benefiting mainstream media outlets over independent YouTube channels. Fox News is likely disproportionally favored on the right due to a lack of other right-leaning mainstream outlets, while traffic in the Center/Left MSM and Partisan Left is more evenly distributed among their representative mainstream outlets.

\begin{figure}[ht]
\centering
    \includegraphics[width=\linewidth]{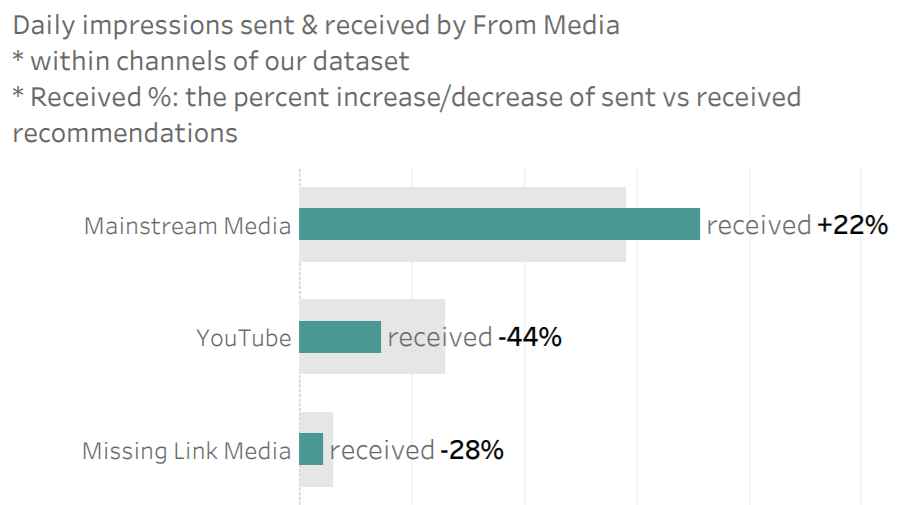}
    \caption{Algorithmic Advantage of Mainstream Media}
    \label{msm}
\end{figure}

We can also analyze the overall net benefits the mainstream media channels are receiving from the algorithm by aggregating the mainstream channels into one high-level group and independent YouTubers into another group and comparing the algorithmic advantages and disadvantages for each. The third group we separated from mainstream media and YouTubers is the group we called the "Missing Link Media." This group encompasses media outlets that have financial backing with the traditional mainstream outlets but are not considered part of the conventional mainstream media. For example, left-wing channels such as Vox or Vice belong to this category, while BlazeTV is an equivalent for the right-leaning media. Figure \ref{msm} shows the clear advantage mainstream media channels receive over both independent channels and Missing Link Media channels.

Finally, based on the findings and analysis of our four claims, we conclude that these data offer little support to the claims that YouTube's recommendation algorithm will recommend content that might be contributing to the radicalization of the user-base. Only the first claim is partially supported, while the data refute all the other three claims. Rejection of these claims seems to be in line with studies that critique the claims of YouTube's algorithm as a pathway to radicalization \cite{munger2019supply}.

\begin{table}[ht]
\caption{Claims and Data Support}
\begin{center}
\begin{tabular}{|p{6cm}|p{2cm}|}
\hline
\textbf{Claim}&{\textbf{Data Support}} \\

\hline
\textbf{C1 - Radical Bubbles.} Recommendations influence viewers of radical content to watch more similar content than they would otherwise, making it less likely that alternative views are presented. & Partially supported \\
\hline
\textbf{C2 - Right-Wing Advantage.} YouTube's recommendation algorithm prefers right-wing content over other perspectives. & Not supported\\
\hline
\textbf{C3 - Radicalization Influence.} YouTube's algorithm influences users by exposing them to more extreme content than they would otherwise. & Not supported\\
\hline
\textbf{C4 - Right-Wing Radicalization Pathway.} YouTube algorithm influences viewers of mainstream and center-left channels by recommending extreme right-wing content, content that aims to disparage left-wing or centrist narratives.  & Not supported  \\
\hline
\end{tabular}
\label{tab1}
\end{center}
\end{table}

YouTube has stated that its algorithm will favor more recent videos that are popular both in terms of views as well as engagement \cite{zhao2019recommending}. The algorithm will recommend more videos based on a user profile, or the most current, popular videos for anonymous viewers. YouTube has stated that they are attempting to maximize the likelihood that a user will enjoy their recommended videos and will remain on the platform for as long as possible. The viewing history determines whether the algorithm will recommend the viewer more extreme content. Antithetical to this claim is that our data show that even if the user is watching very extreme content, their recommendations will be populated with a mixture of extreme and more mainstream content. YouTube is, therefore, more likely to steer people away from extremist content rather than vice versa.

\section{Limitations and Conclusions}

There are several limitations to our study that must be considered for the future. First, the main limitation is the anonymity of the data set and the recommendations. The recommendations the algorithm provided were not based on videos watched over extensive periods. We expect and have anecdotally observed that the recommendation algorithm gets more fine-tuned and context-specific after each video that is watched. However, we currently do not have a way of collecting such information from individual user accounts, but our study shows that the anonymous user is generally directed towards more mainstream content than extreme. Similarly, anecdotal evidence from a personal account shows that YouTube suggests content that is very similar to previously watched videos while also directing traffic into more mainstream channels. That is, contrary to prior claims; the algorithm does not appear to stray into suggesting videos several degrees away from a user's normal viewing habits.

Second, the video categorization of our study is partially subjective. Although we have taken several measures to bring objectivity into the classification and analyzed similarities between each labeler by calculating the intraclass correlation coefficiencies, there is no way to eliminate bias. There is always a possibility for disagreement and ambiguity for categorizations of political content. We, therefore, welcome future suggestions to help us improve our classification.

In conclusion, our study shows that one cannot proclaim that YouTube's algorithm, at the current state, is leading users towards more radical content. There is clearly plenty of content on YouTube that one might view as radicalizing or inflammatory. However, the responsibility of that content is with the content creator and the consumers themselves. Shifting the responsibility for radicalization from users and content creators to YouTube is not supported by our data. The data shows that YouTube does the exact opposite of the radicalization claims. YouTube engineers have said that 70 percent of all views are based on the recommendations \cite{zhao2019recommending}. When combined with this remark with the fact that the algorithm clearly favors mainstream media channels, we believe that it would be fair to state that the majority of the views are directed towards left-leaning mainstream content.  

We agree with the Munger and Phillips (2019), the scrutiny for radicalization should be shined upon the content creators and the demand and supply for radical content, not the YouTube algorithm. On the contrary, the current iteration of the recommendations algorithm is working against the extremists. Nevertheless, YouTube has conducted several deletion sweeps targeting extremist content \cite{martienau2019hate}. These actions might be ill-advised. Deleting extremist channels from YouTube does not reduce the supply for the content \cite{munger2019supply}. These banned content creators migrate to other video hosting more permissible sites. For example, a few channels that were initially included in the Alt-right category of the Ribero et al. (2019) paper, are now gone from YouTube but still exist on alternative platforms such as the BitChute. The danger we see here is that there are no algorithms directing viewers from extremist content towards more centrist materials on these alternative platforms or the Dark Web, making deradicalization efforts more difficult \cite{hussain2014jihad}. We believe that YouTube has the potential to act as a deradicalization force. However, it seems that the company will have to decide first if the platform is meant for independent YouTubers or if it is just another outlet for mainstream media. 

\subsection{The Visualization and Other Resources}

Our data, channel categorization, and data analysis used in this study are all available on GitHub for anyone to see. Please visit the \href{https://github.com/markledwich2/YouTubeNetworks}{GitHub page} for links to data or the \href{https://www.recfluence.net/}{Data visualization}. We welcome comments, feedback, and critique on the channel categorization as well as other methods applied in this study. 

\subsection{Publication Plan}
This paper has been submitted for consideration at \href{https://firstmonday.org}{First Monday}. 

\subsection{Acknowledgments}
First, we would like to thank our volunteer labeler for all the hours spent on YouTube. We would also like to thank Cody Moser, Brenton Milne and Justin Murphy and everyone else who gave their feedback on the early drafts of this paper and aided the editing.  




\bibliographystyle{./bibliography/IEEEtran}
\bibliography{./bibliography/IEEEabrv,./bibliography/IEEEexample}



%
%
%
\clearpage

\begin{appendices}

\section{Channel Categorization}
\label{categories}

\subsection{Channel Views and Formulas}
\label{formulas}

\noindent We have used several formulas in order to capture the flow of recommendations. The main concept in our study is the \textbf{impression}. Impression is an estimate for the number of times a viewer was presented with a recommendation. We count each of the top 10 recommendations for a video as an "impression". Only YouTube knows true impressions, so we use the following process create an estimate:

\begin{table}[ht]
\begin{tabular}{|p{2cm}|p{6cm}|}
\hline
\textbf{Tag} & \textbf{Examples}\\
\hline
Impressions & An estimate for the number of times a viewer was presented with a recommendation. I.e. we count each of the top 10 recommendations for a video as an "impression". Only YouTube knows true impressions, so we use the following process create an estimate: Consider each combination of videos (e.g. Video A to Video B)\newline
\textit{(A to B impressions) = (recommendations from A to B) / (total recommendations from Video A) x (*A's views) x (recommendations per video = 10)}\\
\hline
Relevant impressions & 
\textit{(A channel's relevance \%) x impressions} \\
\hline
Channel views & The total number of video views since first of January 2018\\
\hline
Daily channel views & \textit{(channel views) * (days in the period videos have been recorded for the channel)}\\
\hline
Relevant channel views & \textit{(daily channel views) * (channel relevance \%)}\\
\hline  
\end{tabular}
\end{table}

\subsection{Tag Aggregation}
\label{aggregation}

In order to create meaningful ideological categories, we have aggregated the tags assigned for each channel. In order to calculate the majority view, each soft tag is assessed independently. For each tag, the number of the reviewer with that rag must tally to more than half. Eighteen categories of soft tags, the soft tags defining left, center, and right, and the hard tags defining the media type, were aggregated for the visualization and data analysis. The following list informs which tags or tag combinations were aggregated to represent an ideology, rather than just a collection of tags. 

\begin{itemize}
\item White Identitarian  \(\rightarrow\)  White Identitarian
\item MRA  \(\rightarrow\)  MRA
\item Conspiracy  \(\rightarrow\)  Conspiracy
\item Libertarian  \(\rightarrow\)  Libertarian
\item AntiSJW and either Provocateur or PartisanRight  \(\rightarrow\)  Provocative Anti-SJW
\item AntiSJW  \(\rightarrow\)  Anti-SJW\footnote{This group has a significant overlap with the intellectual dark web-group as described by Ribero et al. (2019), Munger and Phillips (2019)}
\item Socialist  \(\rightarrow\)  Socialist
\item ReligiousConservative \(\rightarrow\)  Religious Conservative
\item Social Justice or Anti-Whiteness \(\rightarrow\) Social Justice
\item Left or Center 'hard' tag and Mainstream News or Missing Link Media 'hard' tag \(\rightarrow\) Center/Left MSM
\item PartisanLeft  \(\rightarrow\) Partisan Left
\item PartisanRight  \(\rightarrow\) Partisan Right
\item AntiTheist  \(\rightarrow\)  Anti-Theist
\item Everything else \(\rightarrow\) Unclassified
\end{itemize}

\subsection{Hard Tags}
\label{hard}
Hard tags are tags sources from external sources. Any combination of the following tags can be applied to a channel. Hard tags are for comparison between the categorization presented in this paper and other work, academic or otherwise, and also used to distinguish between YouTubers and TV or other mainstream media content. 

\begin{table}[htbp]
\begin{tabular}{|p{6cm}|p{2cm}|}
\hline
\textbf{Tag} & \textbf{Examples}\\
\hline
\textbf{Mainstream News}
Reporting on newly received or noteworthy information. Widely accepted and self-identified as news (even if mostly opinion). Appears in either https://www.adfontesmedia.com or https://mediabiasfactcheck.com.

To tag they should have >\ 30\% focus on politics \& culture. 
&
Fox News, Buzzfeed News\\
\hline
\textbf{TV}
Content originally created for broadcast TV or cable
&
CNN, Vice\\
\hline
Ribeiro et al.'s alt-lite, alt-right, IDW
& As listed in Auditing Radicalization Pathways on YouTube \cite{ribeiro2019auditing}
\\
\hline
\end{tabular}
\end{table}

\subsection{Soft Tags}
\label{soft}
Soft tags are a natural category for US YouTube content. Many traditional ways of dividing politics are not natural categories that would accurately describe the politics of YouTube channels. In general, YouTubers are providing reaction and sensemaking on other channels or current events in the United States. We have created a list of categories that attempt to align the stands taken by the channels more naturally, expanding the categorization beyond the left, center, and right categories. 
 
The tag needs to be engaging in some way to the current meta-discussion about YouTube's influence on politics. Our list of categories intends to cover major cultural topics and label channels to the best of our abilities. We have tried to find specific positions that could be mixed and aggregate in order to create categories that would represent ideologies. 

Our guiding principle is that, in order to apply one of these tags, one should be able to judge the channel by the channel content itself. It is important not to rely on an outside judgment about the channel's content. It is also important to interpret the content with full context: there should be no mind-reading and no relying on a judgment from other sources. There should also be enough channels per each category. If the category is too niche, it should be excluded, unless it is essential for the radicalization pathway theory.



\begin{table}[H]
\begin{tabular}{|p{13cm}|p{4.5cm}|}
\hline
\textbf{Tag} & \textbf{Examples} \\
\hline
\textbf{Conspiracy}
A channel that regularly promotes a variety of conspiracy theories. A conspiracy theory explains an event/circumstance as the result of a secret plot that is not widely accepted to be true (even though sometimes it is).

Example conspiracy theories:
\begin{itemize}
    \item Moon landings were faked
    \item QAnon \& Pizzagate
    \item Trump colluding with Russia to win the election
\end{itemize}
&  
X22Report, The Next News Network\\
\hline
\textbf{Libertarian}
A political philosophy that has liberty as its main principle. Generally skeptical of authority and state power (e.g., regulation, taxes, government programs). Favors free markets and private ownership. 

Note:
To tag someone, this should be the primary driver of their politics. 
Does not include libertarian socialists who also are anti-state but are anti-capitalist and promote communal living. 
& 
Reason, John Stossel, The Cato Institute\\
\hline
\textbf{Anti-SJW}
Channel has to have a significant focus on criticizing "Social Justice" (see next category) with a positive view of the marketplace of ideas and discussing controversial topics. 
To tag a channel, this should be a common focus in their content.
& 
Sargon of Akkad, Tim Pool\\
\hline
\textbf{Social Justice}
The channel promotes 
\begin{itemize}
    \item Identity Politics \& Intersectionality – narratives of oppression though the combination of historically oppressed identities: Women, Non-whites, Transgender
    \item Political Correctness – the restriction of ideas and words you can say in polite society.
    \item Social Constructionism – the idea that the differences between individuals and groups are explained entirely by the environment. For example, sex differences are caused by culture, not by biological sex.
    \end{itemize}
The channel content is often in reaction to Anti-SJW or conservative content rather than purely a promotion of social justice ideas.

The supporters of the content creator are active on Reddit in subreddit called r/Breadtube, and the creators often identify with this label. This tag only includes breadtuber's if their content is criticizing anti-SJW's (promoting socialism is its own, separate tag).
& 
Peter Coffin, hbomberguy\\
\hline
\textbf{White Identitarian}
Identifies-with/is-proud-of the superiority of "whites" and Western Civilization. An example of identifying with "western heritage" would be to refer to the Sistine chapel or Bach as "our culture."  

Often will promote
\begin{itemize}
    \item An ethnostate where residence or citizenship would be limited to "whites" OR a type of nationalist that seek to maintain a white national identity (white nationalism)
    \item A historical narrative focused on the "white" lineage and its superiority
    \item Essentialist concepts of racial differences
\end{itemize}
The content creators are very concerned about whites becoming a minority population in the US/Europe (the Great Replacement - theory)
& 
NPIRADIX (Richard Spencer), Stefan Molyneux\\
\hline
\textbf{Educational}
Channel that mainly focuses on education material, of which over 30\% is focused on making sense of culture or politics. 
& 
TED, SoulPancake\\
\hline
\textbf{Late Night Talk shows}
Channel with content presented humorous monologues about the day's news, guest interviews, and comedy sketches.
To tag, they should have over 30\% focus on politics \& culture. 
&
Last Week Tonight, Trevor Noah\\
\hline
\textbf{Partisan Left}
Channel mainly focused on politics and exclusively critical of Republicans. Would agree with this statement: "GOP policies are a threat to the well-being of the country."
&
The Young Turks, CNN\\
\hline
\textbf{Partisan Right}
Channel mainly focused on politics and exclusively critical of Democrats. Must support Trump. Would agree with this statement: "Democratic policies threaten the nation."
&
Fox News, Candace Owens\\
\hline
\textbf{Anti-theist}
The self-identified atheist who is also actively critical of religion. Also called New Atheists or Street Epistemologists. Usually combined with an interest in philosophy.
&
Sam Harris, CosmicSkeptic, Matt Dillahunty\\
\hline
\textbf{Religious Conservative}
A channel with a focus on promoting Christianity or Judaism in the context of politics and culture.
&
Ben Shapiro, PragerU\\
\hline
\textbf{Socialist (Anti-Capitalist)}
Focus on the problems of capitalism. Endorse the view that capitalism is the source of most problems in society.

Critiques of aspects of capitalism that are more specific (i.e., promotion of fee healthcare or a large welfare system or public housing) don't qualify for this tag.

Promotes alternatives to capitalism. Usually, some form of either  Social Anarchist  (stateless egalitarian communities) or Marxist (nationalized production and a way of viewing society through class relations and social conflict). 
&
BadMouseProductions, NonCompete\\
\hline  
\textbf{Revolutionary}
Endorses the overthrow of the current political system. For example, many Marxist and Ethno-nationalists are revolutionaries because they want to overthrow the current system and accept the consequences.
&
Libertarian Socialist Rants, Jason Unruhe\\
\hline
\textbf{Provocateur}
Enjoys offending and receiving any kind of attention (positive or negative). Takes extreme positions, or frequently breaks cultural taboos. Often it is unclear if they are joking or serious.&
StevenCrowder, MILO\\
\hline
\textbf{MRA (Men’s Rights Activist)}
Focus on advocating for rights for men. See men as the oppressed sex and will focus on examples where men are currently oppressed.

Incels, who identify as victims of sex inequality, would also be included in this category.&
Karen Straughan\\
\hline
\textbf{Missing Link Media}
Channels funded by companies or venture capital, but not large enough to be considered "mainstream." 

They are generally accepted as more credible than independent YouTube content.&
Vox, NowThis News\\
\hline
\textbf{State Funded}
Channels that are funded by governments.&
PBS NewsHour, Al Jazeera, RT\\
\hline
\textbf{Anti-Whiteness}
A subset of Social Justice that, in addition to intersectional beliefs about race, has a significant portion of content that essentializes race and disparages "whites" as a group. Channel should match most of the following:
\begin{itemize}
    \item  Negative generalization about "whites". E.g. "White folks are unemotional, they hardly even cry at funerals," e.g., How To Play The Game w/WS 5 Daily Routines
 \item Use of the word "whiteness" as a slur, or an evil force. e.g., "I try to be less white" (Robin DiAngelo)
 \item Simplistic narratives about American history, where the most important story is of slavery and racism.
 \item Dilute terms like racism or white supremacy so that they include most Americans while keeping the stigma and power of the word.
 \item content exclusively framing current events into racial oppression. Usually in the form of police violence against blacks, x-while-black (e.g., swimming while black, walking while black)...
\end{itemize} &
African Diaspora News Channel\\
\hline  
\end{tabular}
\end{table}

\clearpage
\section{Detailed Algorithmic Advantages and Disadvantages}
\label{advantagedisadvantage}

We discuss algorithmic advantages and disadvantages at the higher level in Section \ref{findings}. This appendix presents two additional figures that shows a breakdown of recommendation algorithm traffic channel by channel. 

\begin{figure}[ht]
\centering
    \includegraphics[width=9cm]{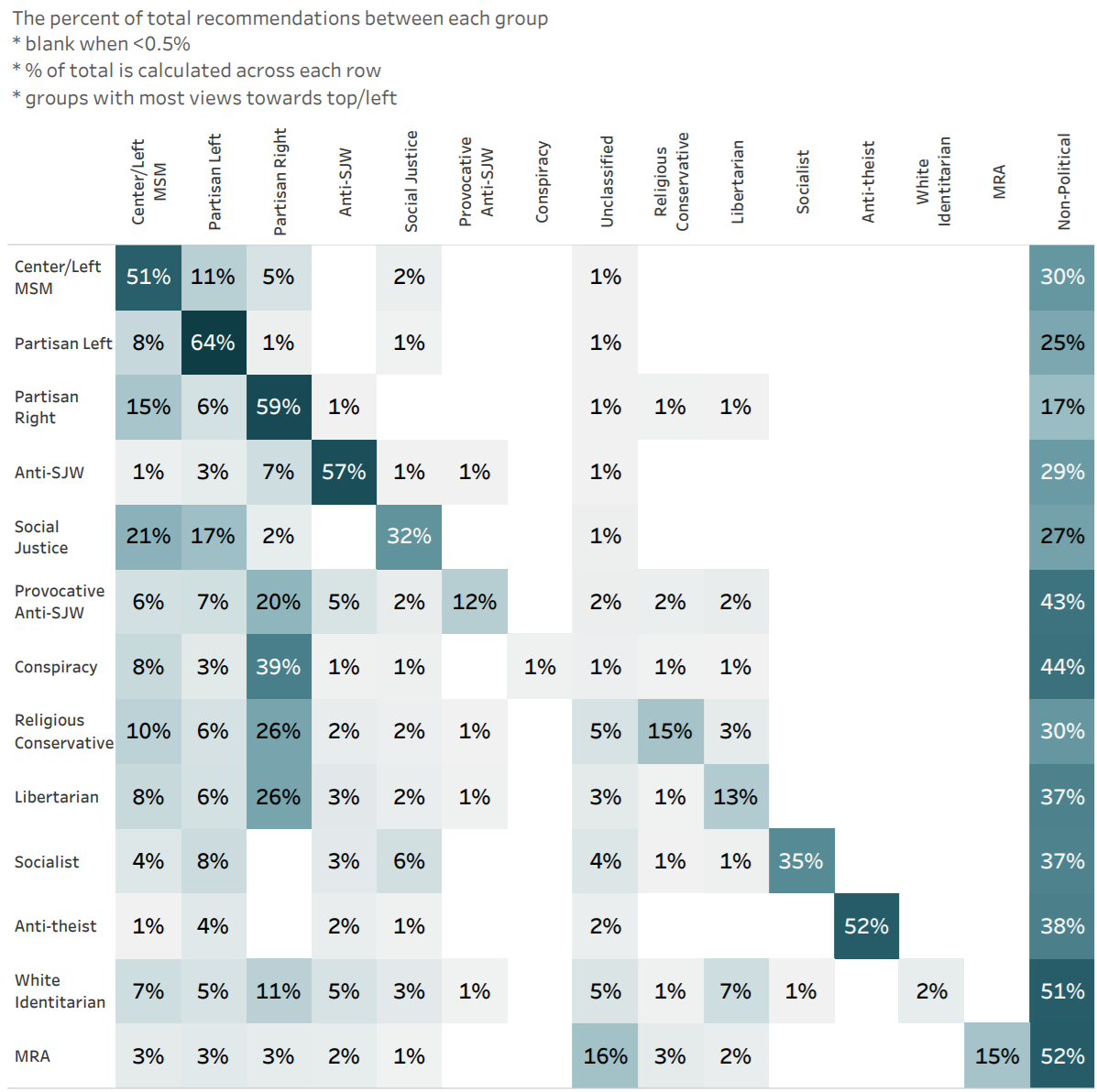}
    \caption{Cross-category and Intra-category Recommendations}
    \label{advantagepercent}
\end{figure}

First, Figure \ref{advantagepercent} presents the relative portion of recommendations between groups. The diagonal column cutting across the chart shows the percentages of intra-category recommendations, i.e., the percentage of recommendations that are directed to the same category. In contrast, lower percentages in this diagonal that the majority of the traffic is directed outwards from the category. The other cells show the percentages each group is recommended in relation to other categories. For example, if one is to view a video that belongs to the Provocative Anti-SJW category, the bulk of the recommendations will suggest videos that belong to either Partisan Right or non-political channels.

The non-political channels in this chart are channels that fall outside our labeled data categories. The Figure \ref{advantagepercent} illustrates that these channels are recommended in large numbers for categories that fall on the fringes, such as the White Identitarian and MRA channels, directing the traffic towards less contentious material.

Figure \ref{advantagepercent} presents the different advantages and disadvantages each group has due to the recommendation system in more detail. The Figure compares the daily net flow of recommendations for each group. The categories in the Figure are organized based on their algorithmic advantage, the most advantaged groups are at the top, and least advantaged groups are at the bottom. Categories in darkest shades of blue are most advantaged, whereas the categories on darker shades of red are at least advantage. 


\begin{figure}[ht]
\centering
    \includegraphics[width=10cm]{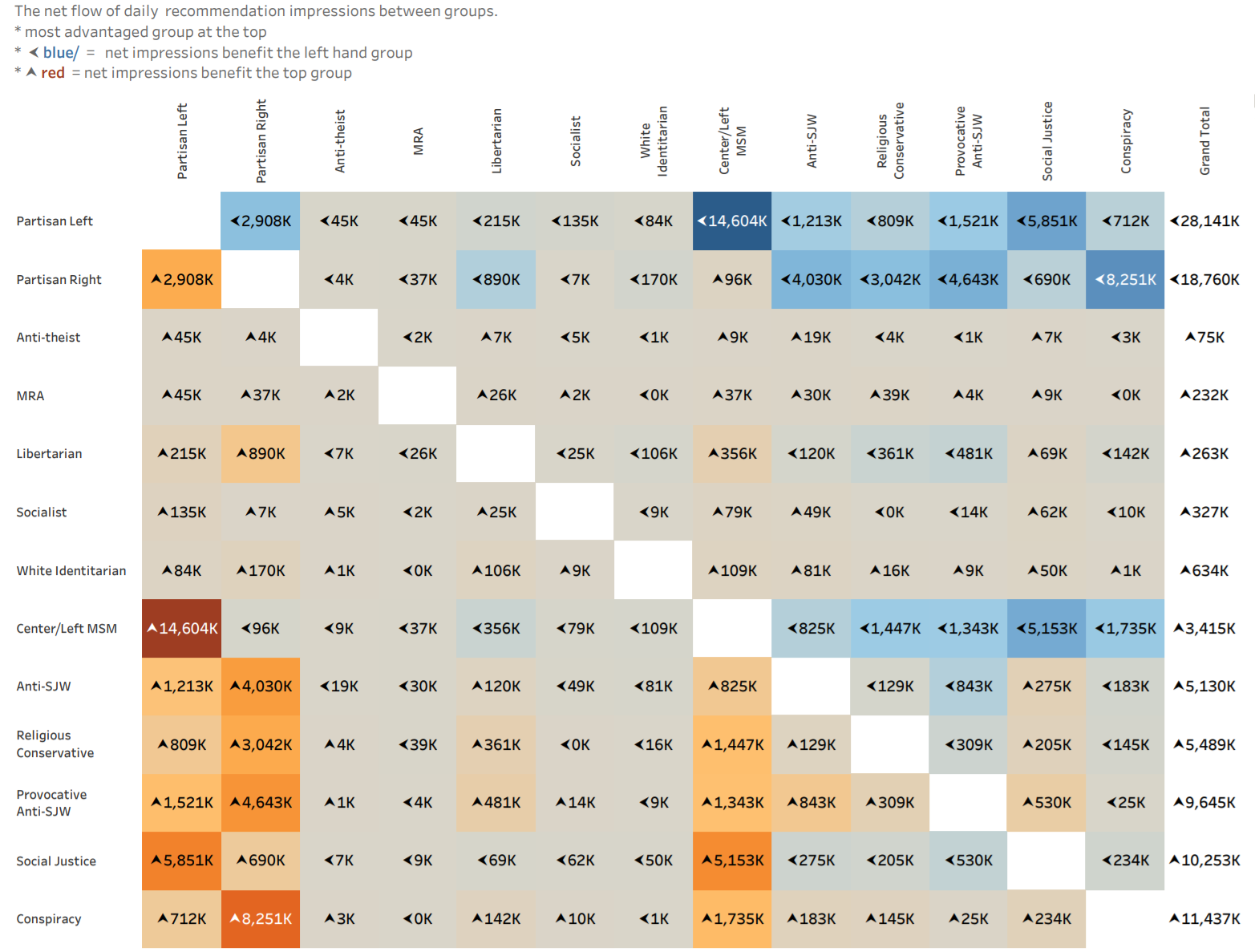}
    \caption{Algorithmic Advantages/Disadvantages in Recommendation Impressions}
    \label{advantagenumber}
\end{figure}

Categories in grey are also at a disadvantage, but to a lesser extent than the categories in red — the small arrows in the image point towards the category, which is benefiting from the recommendations algorithm. Arrows are pointing towards the group that receives more recommendations that it is given by the algorithm, i.e., pointing towards the group, which is advantaged.

...
\end{appendices}
\end{document}